\documentstyle[prl,aps,preprint]{revtex}

\begin{document}
\newcommand{\ket}[1] {\mbox{$ \vert #1 \rangle $}}
\newcommand{\bra}[1] {\mbox{$ \langle #1 \vert $}}
\newcommand{\bkn}[1] {\mbox{$ < #1 > $}}
\newcommand{\bk}[1] {\mbox{$ \langle #1 \rangle $}}
\newcommand{\scal}[2]{\mbox{$ < #1 \vert #2 > $}}
\newcommand{\expect}[3] {\mbox{$ \bra{#1} #2 \ket{#3} $}}
\newcommand{\ki}{\mbox{$ \ket{\psi_i} $}}
\newcommand{\bi}{\mbox{$ \bra{\psi_i} $}}
\newcommand{\p} \prime
\newcommand{\e} \epsilon
\newcommand{\la} \lambda
\newcommand{\om} \omega   \newcommand{\Om} \Omega
\newcommand{\cc}{\mbox{$\cal C $}}
\newcommand{\w} {\hbox{ weak }}
\newcommand{\al} \alpha
\newcommand{\bt} \beta
\newcommand{\be} {\begin{equation}}
\newcommand{\ee} {\end{equation}}
\newcommand{\ba} {\begin{eqnarray}}
\newcommand{\ea} {\end{eqnarray}}
\def\lrD{\mathrel{{\cal D}\kern-1.em\raise1.75ex\hbox{$\leftrightarrow$}}}
\def\lr #1{\mathrel{#1\kern-1.25em\raise1.75ex\hbox{$\leftrightarrow$}}}
\title{ Hawking Radiation from Feynman Diagrams}
\author{R. Parentani
\footnote{e-mail: parenta@celfi.phys.univ-tours.fr}
}
%\vskip 5 truemm
\address{
%\centerline{\em 
Laboratoire de Math\'{e}matiques et Physique
Th\'{e}orique, 
CNRS UPRES A 6083,
%}\centerline{\em 
Universit\'{e} de Tours, 
37200 Tours, France}

\maketitle

\begin{abstract}
%\begin{multicols}{2}
The aim of this letter is to 
clarify the relationships between Hawking radiation
and the 
scattering of light by matter falling into a black hole. 
To this end we analyze the S-matrix elements of a
 model composed of a massive infalling
particle (described by a quantized field) and 
the radiation field. These fields are 
coupled by current-current interactions
and propagate in the Schwarzschild geometry.
As long as the photons energy is much smaller than the 
mass of the infalling particle, one recovers Hawking radiation 
since our S-matrix elements identically reproduce the 
Bogoliubov coefficients obtained by treating the trajectory 
of the infalling particle classically. 
But after a brief period, the energy of the `partners'
of Hawking photons reaches this mass 
and the production of thermal
photons through these interactions stops.
The implications of this result are discussed.
\end{abstract}

\newpage

%\begin{multicols}{2}

\section{Introduction}

Upon deriving black hole radiance, 
Hawking\cite{Hawk} found that $\om$, the
frequency of the in-modes
involved in the Bogoliubov coefficients, 
grows exponentially according to 
\be
\om = \la \; e^{\kappa (u - u_0)}
\label{dopef}
\ee
where $\kappa$ is the surface gravity of the 
 hole and where $u$ is the retarded time around which the
out-particle of energy $\la$ is centered.
This point was emphasized by Gerlach\cite{gerlach}
(and subsequently in \cite{bp}) who
showed that the constant emission rate 
arises from a steady conversion 
of vacuum configurations of frequency  $\om$ 
%characterized by the Doppler shifted frequencies $\om$ 
into red-shifted on shell photons of energy $\la$.

This observation questions the validity of the settings
used to derive Hawking radiation, 
namely 
%linear Bogoliubov transformation 
%obtained from the propagation of a quantum matter 
free field propagation in a given geometry.
Indeed, knowing that gravitational interactions
grow with the energy, what is the validity of
describing these high  frequency configurations
%`transplanckian' frequencies $\om$ 
by free field theory\cite{THooft,unruh,Verl3,MP2}.
%Secondly, can one conceive the origin of the 
%modes, i.e. the degrees of freedom, which are 
%steadily converted into on shell Hawking quanta 
%when the gravitational interactions are no 
%longer neglected. Both aspects have been 
%much discussed, see e.g. \cite{}-\cite{}.
%
% of these questions, 
%we do not know how precisely the 
This question becomes particularly important when one 
considers the interactions between the collapsing matter 
%which forms the hole and 
%the radiation field which leads to
and Hawking quanta. In this respect, 
what is well understood is that
if one first solves Einstein's equations to determine
the collapsing metric 
%driven by the infalling matter
and then study free field propagation in this geometry, 
one obtains Hawking radiation through the frequency 
mixing described in eq. (\ref{dopef}).
In this derivation, the motion (and the quantum state)
of infalling matter is unaffected by the emission of Hawking quanta.
Indeed, the configurations giving rise to these quanta
freely propagate through the collapsing matter\cite{Hawk,MP2}.

The aim of this article is to provide a more dynamical
description of the interactions between infalling matter
and the radiation field $\phi$.
To this end we analyze a simple model
%we shall use is that of \cite{recmir}. It 
 composed of $\psi$, a field 
of mass $m$, coupled to $\phi$ by
%to the radiation massless field $\phi$ through 
current-current interactions and propagating 
in Schwarzschild geometry.
Thus the quanta of $\psi$ represent additional
particles falling into an already formed black hole.
%these 
Since the interactions are described by Feynman diagrams,
%In this new description, 
the infalling particles are now properly
scattered according to energy-momentum conservation.
This is crucial since it reveals the dynamical role 
played by $\om$, the `transplanckian'
energy of the `partner' 
of an asymptotic quantum of energy $\la$. 
Indeed, we now obtain two different 
regimes. First, as long as $\om$ is 
much smaller than $m$, the mass of the infalling 
dust particle, the S-matrix elements of our 
model identically reproduces the Bogoliubov coefficients
obtained by attributing a given inertial trajectory
to the infalling particle. In this regime
one thus recovers the infalling mirror description of 
Hawking radiation\cite{Grov,Carl,Wilc,CV,TV} with 
one important improvement:
the residual energy crossing the future horizon is 
equal to the energy of the infalling particle 
minus the energy carried away
by the Hawking quanta. Therefore, one
does not need
to appeal to Einstein's equations
in order to obtain the notion of black hole evaporation.
%This is the first result of this letter.

The second regime occurs 
when $\om$ becomes comparable to $m$.
Then the scattering amplitudes no longer agree
with those found by Hawking.
In fact, because of energy conservation, 
we shall see that the production of thermal photons 
(induced by scattering on the infalling particle) 
stops when $\om$ reaches $m$. 
With this result, we reach the heart of the problem: 
how to obtain a steady conversion of vacuum 
configurations giving rise to a constant thermal flux 
once recoil effects are no longer neglected.
The questions raised by our negative result are
addressed at the end of the letter. 

It should also be mentioned that other  
derivations of Hawking radiation
do not confront the transplanckian problem in those radical terms. 
First, superspring theory succeeded in deriving
Hawking radiation in completely different settings
%. Indeed in this derivation, Hawking radiation 
since it results
from the degeneracy of black hole microstates\cite{calmad}.
Moreover, the derivation is performed in flat space. 
Thus one confronts neither 
%  transplanckian frequencies and not from 
exponentially growing Doppler effects, 
which are the all mark of regular horizons,
nor therefore the high frequencies.

Secondly, phenomenological models
based on analogies with condensed matter physics
have been proposed
%, see e.g \cite{dumbUnruh,dumbus},
in order to question the relevance of 
these high frequencies\cite{dumbUnruh,dumbus}. 
%Their main weakness however
At present however, one does not know how to
justify dynamically these rather ad hoc models.
%  present lack of understanding to their dynamical justification. 
Perhaps the present approach can provide 
the roots for such a justification.
% for the phenomenological models
%as well as bringing the old approach of Hawking
%%radiation to the approach offered
%by string theory.

%can one conceive the origin of the 
%modes, i.e. the degrees of freedom, which are 
%steadily converted into on shell Hawking quanta 
%when the gravitational interactions are no 
%longer neglected. Both aspects have been 
%much discussed, see e.g. \cite{}-\cite{}

\section{The Model}

In this letter, for simplicity,
% in this letter, 
we shall consider only radial
motion. Therefore, we can work with 2 dimensional
fields. It is then convenient to use 
the conformally flat coordinates $t, z$
in which the 2D line element reads
\ba
ds^2 &=& (1 - {2M \over r}) \left( - dt^2 + dz^{2}
\right) 
\nonumber\\
&=& (1 - {2M \over r})\; \eta_{\mu \nu} dx^\mu dx^\nu
\label{conffl}
\ea
where $\eta^{\mu \nu}$ is the 2D Minkowski unit matrix
and $z$ the tortoise coordinate ($= r + 2M \ln(r/2M -1)$).

In this coordinate system,
the action of the system is 
\ba
S &=&  
%S_\psi + S_\phi + S_{int.}
%\nonumber\\ &=& {1 \over 2}
\int \!dtdz  \left( - \eta^{\mu \nu} 
\partial_\mu \psi^* \partial_\nu \psi
%\vert \partial_t \psi \vert^2
%-  \vert \partial_x \psi  \vert^2
 -(1 - {2M \over r}) m^2\vert \psi  \vert^2
\right) 
\nonumber \\
&&+ \int \!dtdz \left(
\ - \eta^{\mu \nu} 
\partial_\mu \phi^*  \partial_\nu \phi
%\vert  \partial_t \phi \vert^2
%-   \vert \partial_x \phi \vert^2
%\right) \nonumber\\
%&&\ 
- g
%\int \!dtdz \; 
\eta^{\mu \nu} J^\psi_\mu J^\phi_\nu 
\right) 
\label{stot}
\ea
%$\eta^{\mu \nu}$ is the Minkowski unit matrix and 
where 
%$g$ is the coupling constant and 
$J^\psi_\mu = \psi^*\! \lr{i\partial_\mu} \psi $, $J^\phi_\nu 
=\phi^*\! \lr{i\partial_\nu} \phi $ are the currents
carried by the complex fields $\psi$ and $\phi$.
 $g$ is the coupling constant.
We have chosen to work with two complex fields in order to 
have well defined current operators 
before splitting positive and negative frequencies,
see \cite{recmir} for more details concerning this model.

In the absence of interactions ($g=0$), the modes
of the fields freely propagate in Schwarzschild geometry.
Since the metric is static, they can be labeled
by their constant energy. The massless modes
describing infalling and outgoing photons are
respectively
\ba
\phi_\om &=& {e^{- i \om (t + z)} \over \sqrt{4 \pi \om}}
= {e^{- i \om_\mu x^\mu} \over \sqrt{4 \pi \om}}
\nonumber\\
\phi_\la &=& {e^{- i \la (t - z)} \over \sqrt{4 \pi \la}}
= {e^{- i \la_\mu x^\mu} \over \sqrt{4 \pi \la}}
\label{inout}
\ea
In the WKB approximation, the infalling mode of the 
massive field with energy $\e$ is 
\be
\psi_\e=  {e^{- i (\e t + \int^{z} \!dz' p_\e) } 
\over \sqrt{4 \pi p_\e(z)}}
\label{inf}
\ee
where $p_\e(z)$ is the classical momentum, the
positive solution of the mass-shell condition
\be
\eta^{\mu \nu} p_\mu p_\nu = -\e^2 + p^2_\e = - m^2 (1 - 2M/r)
\label{psquare}
\ee
For $m \gg \kappa$ the WKB approximation is valid 
%for $1/m \ll M$
%(i.e. the Compton wave length much smaller than the
%Schwarzschild radius) 
if one does not approach the turning point 
$p_\e(z) =0$ which exists when $\e < m$.

In the interacting picture, to first order in $g$, 
the transition amplitudes are given by the matrix
elements of $g\int\!dtdz J^\psi_\mu J_{\phi}^{\mu}$.
Denoting $A(\e; \om, \la)$ the amplitude 
for an infalling photon of energy $\om$ 
to be scattered by a dust particle of 
energy $\e$ and converted into an 
outgoing photon of energy $\la$,
%To first order in $g$, 
one gets
\ba
A(\e; \om, \la) &=& -i g \int\!dz  
%-\eta^{\mu \nu}
%\left( \om^\mu + \la^\mu \right) 
\left( p_\mu(\e) + p_\mu(\e + \om - \la) \right)
\left( \om^\mu + \la^\mu \right) 
\nonumber\\
&& \quad \times
{e^{-i z(\la + \om)} \over 4 \pi \sqrt{\la \om}}
{ e^{-i \int^{z}\! dz' (p_\e -p_{\e + \om - \la})} 
\over 2 \sqrt{ p_\e \; p_{\e + \om - \la}}}
%{e^{i \om z} \over \sqrt{4 \pi \om}}
\label{A}
\ea
where the final energy of the particle 
is $\e+\om -\la$ since energy conservation
is implemented by the integration over $t$.
The prefactor in the first line of eq. (\ref{A})
 comes from the matrix elements of the two current operators.

By crossing symmetry, 
the amplitude to spontaneously produce these
two photons of energy $\om$ and $\la$ from
scattering by an infalling particle of energy $\e$ is given
by 
\be
B(\e; \om, \la) = A(\e; -\om, \la) 
\label{B}
\ee
As we shall see, it is through
these 
%quite conventional 
Bremsstrahlung-like 
amplitudes that Hawking radiation will be recovered 
in the low energy regime.

\section{Low Energy Regime}

In this regime,
i.e. for $\om, \la \ll m$, one can develop
the phase and the prefactor 
of the integrand of $A(\e; \om, \la)$ in powers 
of $\om$ and $\la$. To first order, we get
\ba
A(\e; \om, \la) &\simeq& -i g \int\!dz 
\left[ (\om + \la)  dt_{cl} /dz 
- (\om - \la ) \right]
\nonumber\\
&&\quad \times
{e^{-i z (\la + \om)} e^{-i t_{cl}(z) (\om - \la)}
\over 4 \pi \sqrt{\la \om} }
%\partial_\e \int^{z'} dz' p_\e}  }
%{e^{i \om z} \over \sqrt{4 \pi \om}}
\label{A2}
\ea
where 
\be
t_{cl}(z) = - \partial_\e \int^{z} \!dz' p_\e(z')
\label{hj}
\ee
is the time lapse evaluated along the infalling
particle trajectory. We also used
$\e/p_\e(z)=dt_{cl}/dz$.

To first order in $\om, \la$, we find that $A(\e; \om, \la)$ 
is proportional to $\alpha(\om, \la)$, the overlap of the photons
wave functions evaluated along the classical trajectory
of the infalling particle characterized by $\e$
and parametrized by $z$ through $t_{cl}(z)$.
%in virtue of Hamilton-Jacobi 
%equations. 
This is easily verified by direct computation.
Similarly, one finds that the amplitude $B(\e; \om, \la)$ 
is proportional, with the {\it same} factor,
to $\beta(\om, \la)= \alpha(-\om, \la)$, the Bogoliubov coefficient
encoding pair creation. 
It should be stressed that 
these agreements to first order in the energy-momentum transfers,
here given by $\om$ and $\la$, 
are generic in character, see \cite{bfa}.

We now focus on the late time regime, for $r/2M-1 
\simeq e^{2 \kappa z} \ll 1$.
In this case one has\cite{gerlach,bp}
\be
{dt_{cl} \over dz} = - 1 -  e^{2\kappa (z - z_0) }
\label{traj}
\ee
Then upon applying the stationary phase condition 
to the integrand of eq. (\ref{A2}) one
recovers eq. (\ref{dopef})
since in the late time regime $z = - u/2 + constant$ 
along the infalling trajectory.
%upon applying the stationary phase condition 
%to the integrand of eq. (\ref{A2}).
This confirms that
\be
\vert{ B(\e; \om, \la) \over A(\e; \om, \la) }\vert^2
= \vert{ \beta(\om, \la) \over  \alpha(\om, \la) }\vert^2
= e^{- 2\pi \la / \kappa}
\label{bovera}
\ee
Eq. (\ref{bovera}) establishes the thermal character
of the outgoing  photons spontaneously emitted 
by the scattering on the infalling particle.
As usual `spontaneously' means that 
one starts from vacuum configurations on ${\cal{I}}^-$.

It is now appropriate to show how the notion of a constant
rate occurs.
% in this saddle point approach. 
To order $g^2$, the mean number of photons of energy $\la$
found on ${\cal{I}}^+$, is given by  
\be
\langle n_\la \rangle =
\int\!d\om  \; \vert B(\e; \om, \la) \vert^2
\label{w}
\ee
To first order in $\om, \la$ it is proportional
to $\int\! d\om /\om$, as in \cite{bp,GO}. To give meaning to this integral,
one uses the fact that at time $u$, the $\la$ photons
arise from frequencies $\om$ centered according to eq. (\ref{dopef}).
This yields $d\om /\om = \kappa du$,
i.e. that the production rate is constant.
Similarly, the mean number of photons received before a certain
time $u$, is obtained by integration over $\om$ up to $\la
e^{\kappa(u - u_0)}$.
%. From this one immediately gets that the mean number 
Thus it increases linearly with $u - u_0$.
%i.e. that the production rate is constant.
%the lapse of proper time 
%on ${\cal{I}}^+$ of this integral.
%is obtained by fixing the upper value 
%to be given by eq. (\ref{dopef}), see \cite{bp,GO}.
This establishes that the thermal flux 
originates from a steady conversion of vacuum
configurations into pairs of photons whose energy
are related by 
%on shell quanta of energy $\la$ through the
%varying Doppler shift
 eq. (\ref{dopef}).

Thus, in this linear approximation in $\om, \la$, one 
recovers
%the steady thermal production of outgoing photons.
Hawking radiation as obtained from free field theory.
Moreover, in our description based on matrix elements,
the {\it residual} energy which crosses
the future horizon is equal to $\e$ minus 
the energy of the outgoing $\la$ quanta
since energy is conserved and since 
the $\om$ quanta fall into the hole
and do not enter into this global energy balance.
Therefore we obtain the 
notion of evaporation without having 
used Einstein's equations, but simply by having
followed the standard rules of quantum
field theory.

The most remarquable
feature of these results is the steadiness of the 
production rate. However, as we shall see,
it is a consequence of having performed 
a first order expansion in $\om$.
Moreover, because of eq. (\ref{dopef}),
$\om$ reaches $m$
after a few e-folds in the units of $1/\kappa$.
Therefore, since the low energy regime is brief,
it is mandatory to take into account higher order effects
in $\om/\e$.

\section{High Energy Regime}

To characterize the high energy regime
we first analyze the classical channel, i.e. the
scattering of an infalling photon of frequency $\om$ 
which is sent from ${\cal{I}}^-$.
The simplest way to proceed
consists in applying the stationary phase condition
to the phase of the integrand of $A(\e; \om, \la)$.
Then, the dominant contribution arises from values
of $z$ centered around the saddle point value $z_{sp}$,
the solution of 
\be
\om + \la = p_{\e +\om -\la}(z_{sp}) - p_\e(z_{sp}) 
\label{stpc}
\ee
In the late time regime and
for $\om \gg \la \simeq \kappa$, it obeys
\be
e^{2 \kappa z_{sp}} = { 4 \la \over \om} {\e^2 \over m^2} 
( 1 + {\om \over \e} )
\label{stpc2}
\ee
As long as $\om /\e \ll 1$ one recovers the low
energy regime since eq. (\ref{stpc2}) 
is equivalent to eq. (\ref{dopef}).

Instead, when $\om$ aproaches $\e$, 
one reaches a new regime:
 the maximal value of $\la$, the energy of the scattered
photon found on ${\cal{I}}^+$, is bounded by 
$m e^{-\kappa (u- u_0)}/4$
no matter how $\om$ big is. Thus one looses the 
illusionary possibility
offered by eq. (\ref{dopef}),
of receiving, on ${\cal{I}}^+$ and at arbitrarily large times,
photons with a given frequency $\la$ by sending from ${\cal{I}}^-$
quanta with sufficiently high frequencies $\om$. 

In loosing this possibility,
 we recover conventional physics: a particle of mass $m$ cannot
properly reflect photons of energy (measured in its rest frame)
higher than its mass, see eq. (62) in \cite{recmir}.
Then, because of the red shift from the scattering locus $z_{sp}$
to ${\cal{I}}^+$, one obtains the announced bound
for $\la$. We believe that a similar conclusion
should also emerge from taking into 
account the deformation of the geometry
induced by a infalling wave of energy  $\hbar \om$. 
In both cases, eq. (\ref{dopef}) would be 
invalidated after time lapses of 
the order of a few $1/\kappa$.

Having clarified the direct channel,
we now turn to spontaneous pair creation
amplitudes. Due to crossing symmetry, 
the stationary phase conditions applied
to $B(\e; \la, \om)$ are given by eqs. (\ref{stpc})
and (\ref{stpc2}) with $\om$ replaced by $-\om$.

As for the amplitude $A$, the new regime
arises when $\om$ approaches $\e$.
To characterize the onset of this regime,
one should determine the first order corrections 
in $\om/\e$.
%, i.e. one must perform a second
%order expansion of the integrand of $B$. 
This is easily achieved by exploiting the following fact.
Besides the exponential $e^{i(\om - \la ) z}$,
the phase of the integrand of $B$ is an anti-symmetric
function in $\om + \la$ when developed around 
the `mean' energy $\bar \e = \e /2 - (\e - \om -\la)/2$.
Therefore, there is no quadratic terms 
in $\om$ when one develops around $\bar \e$.
(Note in passing that this result directly follows
from quantum mechanics which dictates that the 
wave functions of the `heavy' system enter 
transition amplitudes always in products of the form 
 $\psi^*_{\e_{f\!i\!n}} \psi_{\e_{i\!n}}$, see \cite{bfa,area}.)

Thus the improved value of $B(\e; \om, \la)$ is 
given by the first order expression
with $\e$ replaced by $\bar \e$. 
From this one deduces the following. 
Firstly, the first order correction
to Hawking temperature vanishes. 
Indeed, eq. (\ref{bovera}) still obtains since the value
of $\e$ plays no role to first order in $\om$.
Secondly, the emission rate at which these thermal 
quanta are emitted is modified. Indeed, upon computing
their mean number, eq. (\ref{w}), one faces a
new phase space volume. 
We remind the reader that when dealing with 
eq. (\ref{dopef}), one had $d\om/\om = \kappa du$
which guaranteed a constant flux. Instead, 
upon using eq. (\ref{stpc2}), one gets
\be
{ d\om \over \om } \simeq 
\kappa du \left( 1 - { \om \over  \e} \right)
\label{newr}
\ee
which implies a decreasing rate when $\om/\e $ approaches 1. 

It is of little interest to further characterize
this decrease because
%This decreasing rate, obtained perturbatively
%to second order in $\om$, is conforted by the following fact.
%When $\om$ approaches $\e$, 
one encounters an unavoidable barrier:
the final energy of the 
scattered particle ($=\e -\om -\la$)
cannot be negative. Indeed,
if one follows the conventional rules of QFT, 
wave functions with negative energy do not 
correspond to on shell particles and therefore do not appear
in matrix elements at the tree approximation. 
Therefore, the high energy regime 
(whose unbounded character 
was at the origin of the constant rate
in the usual derivation)
is simply not available since the upper bound 
in the integral of eq. (\ref{w}) is given
by $\e -\la$. 
Thus the steadiness of the emission rate
is lost and the total number of quanta 
emitted is bounded.
(Notice that a different result has been obtained
in \cite{TV} from a somehow more kinematical model.)

\section{Conclusions}

Upon analyzing the scattering amplitudes
$A(\e; \om, \la)$ and $B(\e; \om, \la)$
we found two regimes.
% according to the
%importance of $\om/\e$. 
As long as $\om/\e \ll 1$,
it is legitimate to perform a first order expansion in 
$\om/\e$ and one recovers the usual 
Bogoliubov coefficients. In fact this agreement defines 
the physical interpretation of these coefficients:
they are the correct amplitudes as long as 
backreaction effects can be neglected. This is 
why they only depend on $\om, \la$ and $\kappa$.

When $\om$ approaches $\e$, after a 
few $1/\kappa$ time lapses, the thermal production 
stops. Indeed, because of energy conservation,
one can no longer appeal to unbounded frequencies $\om$,
as one does in the absence of backreaction.

In this we have reached our aim: 
to show when how and why
a simple approach based on QFT fails
to reproduce Hawking radiation at large times.
Needless to say that
it is over hasty to deduce from this 
that Hawking radiation actually stops. Rather
it poses with greater accuracy
the question: which fundamental theory is Hawking's 
derivation an approximation of ?

A first radical option consists in postulating that
Hawking radiation {\it cannot} be recovered from QFT
based on General Relativity once backreaction
effects are included. Instead it 
should emerge from other settings
such as string theory\cite{calmad}. 
Then, our negative result
can be considered as an indication in 
favour of this belief.

A less radical approach consists in hoping
that one shall recover Hawking radiation 
by improving the present analysis. 
(Notice that this 
conservative approach does work for the Unruh 
effect\cite{Unr}. In that case, the backreaction effects obtained
from QFT are bounded and the corrections to the 
Unruh effect stay finite\cite{rec}.)

The improvements 
can be pursued in two directions.
Firstly, by including scattering on many 
dust particles, one might get collective effects
which significantly differ from what we obtained.  
Secondly, by including loop corrections
one could recover access to the infinite reservoir 
of high frequencies and find that Hawking 
radiation does not emerge from on-shell diagrams. 
Indeed, one should explore the consequences of modifying
Feynman rules
for the external legs which correspond to
quanta falling into the hole
since there is no a priori reason to treat
asymptotic and infalling quanta on the same footings.
In this respect one can already notice that, due to the 
$r$-dependence in eq. (\ref{psquare}), the final energy
of the scattered particle can be
smaller than $m$ even though
the corresponding particle cannot be found asymptotically.

%\end{multicols}

\end{document}